\documentclass[nofootinbib,prd]{revtex4}

\usepackage{graphicx}
\usepackage{dcolumn}
\usepackage{bm}

\begin{document}
\def\mib#1{\mbox{\boldmath $#1$}}
\preprint{OU-HET 541}

\title{Radiative Corrections to the Neutrino-Deuteron Reactions}
\author{Masataka Fukugita}
\email{fukugita@icrr.u-tokyo.ac.jp}
\affiliation{%
Institute for Cosmic Ray Research, University of Tokyo, 
Kashiwa 277-8582, Japan
}%

\author{Takahiro Kubota}
\email{ kubota@het.phys.sci.osaka-u.ac.jp}
\affiliation{
Graduate School of Science, Osaka University, Toyonaka, 
Osaka 560-0043, Japan
}%

\date{\today}

\begin{abstract}
The $O(\alpha)$ QED and electroweak radiative corrections to
neutrino deuteron reactions is investigated with particular
emphasis given to the constant terms, which have not been treated
properly in the literature. This problem is related to the definition
of the axial-vector coupling constant $g_{A}$ as to the inclusion of
radiative corrections.
After proper calculations of the constants for the Fermi and
Gamow-Teller transitions, we find the radiative correction
to the neutral current induced reaction, with the usually adopted
definition  of  $g_{A}$, is 1.017 for the Higgs boson mass
$m_{H}=1.5 \; m_{Z}$. This value is close to that given by
Kurylov et al., but this is due to an accidental cancellation
of the errors, between those caused by putative identification of
constant terms for
the Fermi and Gamow-Teller transitions for the charged current
reactions and minor errors
in their treatment of the constant terms for the
neutral current induced reactions.
\end{abstract}

\maketitle

\section{\label{Intro}Introduction}

We consider radiative corrections to neutrino reactions off
deuterons,
\begin{eqnarray}
& &\nu _{e}+d\rightarrow e^{-} + p + p,
\label{eq:nud}
\\
& &\nu _{e}+d\rightarrow \nu _{e}+ p + n.
\label{eq:nud2}
\end{eqnarray}
These reactions have been used for the solar neutrino measurement
at Sudbury Neutrino
Observatory (SNO), with the accuracy now reached the
level that radiative corrections are non-negligible \cite{sno}.
This measurement has played a crucial role to fully resolve
the long-standing solar neutrino problem \cite{bahcall}.

The first attempt to calculate radiative corrections for these
processes was made by
Towner \cite{towner}. Some subtleties associated with the integral
of the soft photon emission, as questioned in \cite{beacom}, was
remedied by Kurylov et al. \cite{kurylov}, with a careful treatment
of the energy-dependence of the wave function overlap between the
initial and final states. As remarked by the latter authors, however,
there still remains the problem as to the constant terms of
radiative corrections. These authors evaluated the corrections 
for the charged current process by implicitly 
assuming that the inner correction constant to the Gamow-Teller
part is the same as that to the Fermi transition.

The $O(\alpha)$ radiative correction for the charged current
induced reactions is generally written as
\begin{eqnarray}
A(\beta) =\left (1+\delta_{\rm out}(\beta) \right )\left[
f_{V}^{2} \left (1+{\delta_{\rm in}^{\rm F}}
\right ) \langle 1 \rangle ^{2} +
g_{A}^{2} \left (1+
{\delta_{\rm in}^{\rm GT}}\right )\langle {\mib \sigma} \rangle ^{2}
\right]\ ,
\label{eq:Aterm}
\end{eqnarray}
where $ \langle 1 \rangle$ and $\langle {\mib \sigma} \rangle$ are
the Fermi and Gamow-Teller matrix elements, $f_V$ ($=1$)
and $g_A$ are vector and axial-vector coupling constants;
radiative corrections are divided into the outer part $\delta_{\rm out}$
that depends on electron velocity $\beta$, hence is process dependent,
and the inner parts $\delta_{\rm in}^{\rm F}$ and
$ \delta_{\rm in}^{\rm GT}$,
which are universal, irrespective of the process
considered \cite{sirlin}.
The outer correction is common to both Fermi and
Gamow-Teller parts, and
thus factored out.
Specifically for the reaction (\ref{eq:nud}), only the Gamow-Teller
part contributes.
The radiative correction to the neutral current reaction for the
axial-vector induced reaction (\ref{eq:nud2}) is due to electroweak
interactions and is written 
\begin{equation}
B(\beta)= (1+ \Delta_{\rm in}^{\rm GT})g_{A}^{2} \langle {\mib \sigma}
\rangle ^{2} \ .
\label{eq:corrNC}
\end{equation}

Among the radiative correction factors, 
$\delta_{\rm out}(\beta ) \equiv (\alpha /2\pi ) g(\beta )$ has
been known from early times \cite{ks}
and $\delta_{\rm in}^{\rm F}$ was calculated
by Marciano and Sirlin \cite{ms1}. The correction for the Gamow-Teller
transition $ \delta_{\rm in}^{\rm GT}$ was calculated only recently
\cite{fukugita1}. In the absence of the calculation of
$ \delta_{\rm in}^{\rm GT}$, the axial-vector coupling constant
$g_{A}$ is extracted from neutron beta decay using
\begin{eqnarray}
A(\beta) =\left (1+\delta_{\rm out}(\beta )\right )\left (1+{\delta_{\rm
in}^{\rm F}}
\right ) \left[ \langle 1 \rangle ^{2}
f_{V}^{2}
+\langle {\mib \sigma} \rangle ^{2}
\tilde g_{A}^{2} \right]\ .
\label{eq:Aterm1}
\end{eqnarray}
Therefore, the axial-vector coupling usually quoted in the literature is
$\tilde g_{A}$ and not the bare $g_{A}$, nor 
$(1+\delta_{\rm in}^{\rm GT})g_{A}$
which is the form with proper inclusion of the radiative correction.
This does not cause practical problems, however, in so
far as one deals only
with charged current processes, since the inner correction
constant is universal \cite{sirlin}. This is also true for
the process involving polarisation, from which the axial-vector
coupling constant is
extracted \cite{fukugita2}. The extracted $\tilde g_{A}$
differs from $g_{A}$ that appears in Lagrangian by
\begin{eqnarray}
g_{A}^{2}= \frac{1+{\delta_{\rm in}^{\rm F}}}
{1+{\delta_{\rm in}^{\rm GT}}}~ \tilde g_{A}^{2}\ ,
\label{eq:pseudoga}
\end{eqnarray}
so that the effect is absorbed into the
redefinition of the axial-vector coupling constant.

This does not apply, however, to the correction for the neutral
current process.
The radiative correction to the neutral current Gamow-Teller reaction
$\Delta_{\rm in}^{\rm GT}$ can be obtained from the general expression
given by Marciano and Sirlin \cite{ms2}. To unfold $g_A$, however, we need
the knowlege of $ \delta_{\rm in}^{\rm GT}$. In the work of
Kurylov et al. \cite{kurylov} $g_A=\tilde g_A$ is assumed.
The purpose of this paper is to provide
complete constant terms of the $O(\alpha)$ radiative corrections
for both charged and neutral current induced neutrino deuteron
reactions, (\ref{eq:nud}) and (\ref{eq:nud2}).


\section{The constant term of radiative corrections }

The inner radiative corrections that appear in (\ref{eq:Aterm}) are
given by
\begin{eqnarray}
\delta _{\rm in}^{\rm F}&=&\frac{e^{2}}{8\pi ^{2}}\left \{
\frac{3}{2}{\rm log}\left ( \frac{m_{Z}^{2}}{m_{p}^{2}}\right )
+3\bar Q {\rm log}
\left ( \frac{m_{Z}^{2}}{M^{2}} \right ) +C^{\rm F}
\right \},
\label{eq:deltaf}
\\
\delta _{\rm in}^{\rm GT}&=&\frac{e^{2}}{8\pi ^{2}}\left \{
\frac{3}{2}{\rm log}\left ( \frac{m_{Z}^{2}}{m_{p}^{2}}\right )
+1+3\bar Q {\rm log}
\left ( \frac{m_{Z}^{2}}{M^{2}} \right ) +C^{\rm GT}
\right \}\ ,
\label{eq:deltagt}
\end{eqnarray}
where $m_{p}$ and $m_{Z}$ are the
proton and $Z$ boson masses,
and $M$ is the lower energy cutoff that represents the scale
of the onset of asymptotic behaviour of the electroweak
theory and is taken to be of the order of 1 GeV.
The terms that contain $\bar Q$ (=1/6 for the standard quark model)
and the constant terms
$C^{\rm F}$ and $C^{\rm GT}$ depend on the structure
of hadrons and hence are model dependent.
The expression (\ref{eq:deltaf}) and its numerical evaluation
were given in \cite{ms1} and those for
(\ref{eq:deltagt}) were obtained in \cite{fukugita1} in a manner
parallel to \cite{ms1}. The constant terms were evaluated as
\begin{eqnarray}
C^{\rm F}=2.160, \hskip1cm C^{\rm GT}=3.281.
\end{eqnarray}
With these numbers we find
for $M \approx 1 {\rm GeV}$,
\begin{eqnarray}
\delta _{\rm in}^{\rm F}=0.0237, \hskip1cm \delta _{\rm in}^{\rm GT}=
0.0262,
\label{eq:delta}
\end{eqnarray}
where a dominant contribution to $\delta _{\rm in}^{\rm GT}$ comes from
weak magnetism \cite{fukugita1}.

The electroweak radiative corrections to the hadronic matrix elements
of the neutral current
have been worked out in \cite{ms2},
\begin{eqnarray}
{\cal M}_{\rm eff}^{\mu }
&=&
\frac{2im_{Z}^{2}}{q^{2}-m_{Z}^{2}}\frac{G_{\mu }}{\sqrt{2}}\:
\rho _{NC}^{(\nu ; h)} (q^{2})
\langle f \vert \left \{ \bar \psi I_{3}\gamma ^{\mu}
\frac{1-\gamma ^{5}}{2} \psi
- \kappa ^{(\nu ; h)}(q^{2}){\rm sin}^{2}\theta _{W} \bar \psi \gamma ^{\mu}
Q\psi \right \} \vert i \rangle
\nonumber \\
& & + \frac{i(g^{2}+{g'}^{2})}{q^{2}-m_{Z}^{2}}\frac{e^{2}}
{32\pi ^{2} }\frac{1}{{\rm sin}^{2}\theta _{W}{\rm cos}^{2}\theta _{W}}
\langle f \vert \left \{ a_{\beta _{L}}J^{\mu}_{\beta _{L}}+a_{\beta _{R}}
J^{\mu}_{\beta _{R}} \right \} \vert i \rangle ,
\label{eq:marcianosirlin}
\end{eqnarray}
where $I_{3}$ and $Q$ are isospin and electric charge, and
$\rho _{NC}^{(\nu ; h)} (q^{2})$ and $\kappa ^{(\nu ; h)}(q^{2}) $
include the $O(\alpha)$ electroweak corrections; the momentum transfer
$q$ is set equal to zero for our purpose.
The last two terms, induced as $O(\alpha)$ corrections, are 
given by eq. (19) of ref.
\cite{ms2}, but we only need to know that 
the interference between the first two and the last two terms takes
the form
 \begin{eqnarray}
& &
\hskip-1cm \langle u_{L} \vert J_{Z}^{\mu} \vert u_{L} \rangle \langle
u_{L}
\vert J_{\beta _{L}}^{\nu} \vert u_{L} \rangle
+
\langle d_{L} \vert J_{Z}^{\mu} \vert d_{L} \rangle \langle d_{L}
\vert J_{\beta _{L}}^{\nu} \vert d_{L} \rangle
\nonumber \\
& &=\frac{1}{2}\left ( 1-\frac{4}{3} {\rm sin}^{2}\theta _{W} \right )
\left ( 1-\frac{2}{3} {\rm sin}^{2}\theta _{W} \right )
\left (
\bar u_{L}\gamma ^{\mu}u_{L}\cdot \bar u_{L}\gamma ^{\nu}u_{L}
-
\bar d_{L}\gamma ^{\mu}d_{L}\cdot \bar d_{L}\gamma ^{\nu}d_{L}
\right ),
\\
& &
\hskip-1cm
\langle u_{R} \vert J_{Z}^{\mu} \vert u_{R} \rangle \langle u_{R}
\vert J_{\beta _{R}}^{\nu} \vert u_{R} \rangle
+
\langle d_{R} \vert J_{Z}^{\mu} \vert d_{R} \rangle \langle d_{R}
\vert J_{\beta _{R}}^{\nu} \vert d_{R} \rangle
\nonumber \\
& &=
\frac{2}{9}\;{\rm sin}^{2}\theta _{W}
\left (
\bar u_{R}\gamma ^{\mu}u_{R}\cdot \bar u_{R}\gamma ^{\nu}u_{R}
-
\bar d_{R}\gamma ^{\mu}d_{R}\cdot \bar d_{R}\gamma ^{\nu}d_{R}
\right ),
\end{eqnarray}
when matrix elements are evaluated with left- and right-handed
nonstrange quarks. This vanishes for isosiglet targets, so that
the last two terms do not contribute for deuteron reactions. 
For deuterons only the axial current among the
first two term of (\ref{eq:marcianosirlin})
contributes, so that the electroweak radiative correction
gives rise to the axial
coupling $g_{A}$ renormalised as
\begin{eqnarray}
g_{A} \rightarrow \rho _{NC}^{(\nu ; h)}(0)g_{A}.
\label{eq:renormalisedncga}
\end{eqnarray}
This means the correction of
(\ref{eq:corrNC}) 
\footnote{ 
The expression given by
Kurylov et al. \cite{kurylov} retains some contributions  
from the last two terms of (\ref{eq:marcianosirlin}).
It is obvious from symmetry that these terms ought to 
vanish for deuterons.
} 
\begin{eqnarray}
1+\Delta _{\rm in}^{\rm GT}=\rho _{NC}^{(\nu ; h)}(0)^{2}.
\end{eqnarray}

The calculation of Marciano and Sirlin \cite{ms2} gives
\begin{eqnarray}
\Delta _{\rm in}^{\rm GT}&=&\frac{e^{2}}{8\pi ^{2}}\Bigg \{
\frac{3{\rm log}({\rm cos}^{2}\theta _{W})}{4{\rm sin}^{4}\theta _{W}}
-\frac{7}{4{\rm sin}^{2}\theta _{W}}+\frac{2a_{Z}}{{\rm sin}^{2}\theta _{W}
{\rm cos}^{2}\theta _{W}}
+G(\xi ^{2} , {\rm cos}^{2}\theta _{W})+\frac{3}{4{\rm sin}^{2}\theta _{W}}
\cdot \frac{m_{t}^{2}}{m_{W}^{2}}\Bigg \},
\end{eqnarray}
where $m_{t}$ is the top quark mass, 
$\sin ^{2} \theta_W$ is the weak mixing angle in the on-shell scheme,
and
\begin{eqnarray}
a_{Z}&=&\frac{1}{2{\rm cos}^{2}\theta _{W}}\Bigg [
\frac{5}{2}-\frac{15}{4}{\rm sin}^{2}\theta _{W}-\frac{1}{5}{\rm sin}^{4}
\theta _{W} +\frac{14}{9}{\rm sin}^{6}\theta _{W}\Bigg ]
\\
G(\xi ^{2} , {\rm cos}^{2}\theta _{W})
&=&\frac{3\xi ^{2} }{4 {\rm sin}^{2}\theta _{W}}
\Bigg \{ \frac{{\rm log}({\rm cos}^{2}\theta _{W}/\xi ^{2} )}
{{\rm cos}^{2}\theta _{W}
-\xi ^{2} }+\frac{1}{{\rm cos}^{2}\theta _{W}}\cdot \frac{{\rm log}\xi ^{2}
}{1-\xi ^{2}}
\Bigg \},
\end{eqnarray}
with $\xi =m_{H}/m_{Z}$ , $m_{H}$ being the
Higgs boson mass. Numerically, the $\rho _{NC}^{(\nu ; h)}(0)-1$
factor is represented by
\begin{eqnarray}
\rho _{NC}^{(\nu ; h)}(0)-1 &=& 0.010164-0.0004628~\xi+
3.708\times10^{-4}\xi ^{2}
-1.332\times10^{-6}\xi ^{3} \cr
&&
+0.00960\left \{ \left ( \frac{m_t}{ 178\; {\rm GeV}}\right )^{2}
-1 \right \},
\end{eqnarray}
which is correct with the error up to 0.06\% for the range
$1 < m_H/m_Z < 10$.

The radiative-corrected cross section of (\ref{eq:nud2}) is
given by multiplying $1+\Delta _{\rm in}^{\rm GT}$
on the tree value. For $m_t=178 \: {\rm GeV}$, 
\begin{eqnarray}
\rho _{NC}^{(\nu ; h)}=1.00955,
\hskip0.5cm
\Delta _{\rm in}^{\rm GT}&=&0.0192 \hskip1cm {\rm for} \:\: m_{H}=1.5 \:
m_{Z}, 
\\
\rho _{NC}^{(\nu ; h)}=1.00862,
\hskip0.5cm
\Delta _{\rm in}^{\rm GT}&=&0.0173 \hskip1cm {\rm for} \:\: m_{H}=5.0 \:
m_{Z}.
\end{eqnarray}

This, together with (\ref{eq:deltaf}), (\ref{eq:deltagt}), and
(\ref{eq:delta}), gives the complete set of the constants for
the $O(\alpha)$ radiative corrections to the neutrino deuteron
reactions.
In the usual applications, however, the axial coupling used is
$\tilde g_A$ derived using (\ref{eq:pseudoga})
rather than $g_A$ ($g_A=0.9988\tilde g_A)$.
With the use of $\tilde g_A$, the
cross section for the neutral current induced reaction
receives the extra factor $ (1+\delta _{\rm in}^{\rm F})/
(1+\delta _{\rm in}^{\rm GT})$, so that the
correction factor for the cross section reads
\begin{eqnarray}
\left (1+\Delta _{\rm in}^{\rm GT} \right ) \left (
\frac{1+\delta _{\rm in}^{\rm F}}{1+\delta _{\rm in}^{\rm GT}}\right )
=1.017\ ,
\end{eqnarray}
for example, for $m_H=1.5m_Z$. (This value will be 1.015 for
$m_H=5m_Z$).

This number happens to be close to that given by Kurylov et al.
\cite{kurylov}, but it is due to an accidental compensation of the
error arising from a neglect of the difference between
$ \delta _{\rm in}^{\rm GT}$
and $ \delta _{\rm in}^{\rm F}$ by their incorrect treatment of the
constant term in the radiative correction to the neutral current
induced reaction (see footnote above).
The results of the SNO experiment \cite{sno}
using \cite{kurylov}, therefore, remain virtually unchanged.

We note as a final remark that the constant term for
the radiative correction
to the ratio of neutral to charged current reaction (after the usual outer
correction \cite{kurylov, ks} for the charged 
current reaction ) is $-0.6$\%, which may
be compared with the claimed error (0.5\%) of nuclear calculations for the
ratio of tree level cross sections \cite{nakamura}.


\begin{acknowledgments}
We thank Yasuo Takeuchi for useful correspondences.
MF would like to express his sincere thanks to late John Bahcall
for many discussions on neutrino
physics over many years at the Institute for Advanced Study in Princeton.
This work is supported in part by Grants in Aid
of the Ministry of Education.
\end{acknowledgments}



\end{document}